\documentclass[aps,12pt,prl]{revtex4-2}

\usepackage{bm}

\usepackage{natbib}

\usepackage{hyperref} 
\hypersetup{
    colorlinks = true,
    linkcolor={gray},
    citecolor={blue!50!black},
    urlcolor={blue!50!black}
}
\usepackage{siunitx}
\usepackage{mathtools}
\usepackage{bbold}
\usepackage[normalem]{ulem}
\usepackage{enumerate}
\usepackage[dvipsnames]{xcolor}
\def\be{\begin{equation}}
\def\ee{\end{equation}}
\def\e#1{\label{#1}\end{equation}}
\def\bea{\begin{eqnarray}}
\def\eea{\end{eqnarray}}
\def\ea#1{\label{#1}\end{eqnarray}}

\def\bem#1{\begin{mathletters}\label{#1}}
\def\eml{\end{mathletters}}

\def\4#1{{\boldsymbol{#1}}}
\def\8#1{{\widetilde{#1}}}
\def\bse{\begin{subequations}}
\def\ese{\end{subequations}}

\begin{document}

\title{Matters Arising: Time-reversal-based quantum metrology with many-body entangled states}

\author{Liam P. McGuinness}
\affiliation{Laser Physics Centre, Research School of Physics, Australian National University, Acton, Australian Capital Territory 2601, Australia \\ \textnormal{Email: \href{mailto:liam@grtoet.com}{liam@grtoet.com}} }

\begin{abstract}
In their paper ``Time-reversal-based quantum metrology with many-body entangled states'' Nature Physics (2022) \cite{Colombo2022}, Colombo et.~al. claim to measure both an unknown phase and an oscillating magnetic field with a precision that cannot be achieved using independent particles -- a limit known as the standard quantum limit. By entangling an ensemble of $\sim300$ atoms, Colombo et. al. measure an angle of rotation away from a known initial state and additionally measure a magnetic field oscillating at 290\,Hz. The authors report an experimental precision approximately a factor of 4 beyond what is possible with the same number of independent atoms (12.8\,dB and 11.8\,dB for these tasks respectively). These claims are incorrect. Colombo et.~al. do not surpass the precision bound for 300 independent particles, nor do they even surpass the precision bound for a single particle. Colombo et.~al. cite several experiments that surpass the standard quantum limit using entanglement. Each and every paper cited performs incorrect, incomplete or misleading comparisons of the type that we highlight here. The consequence being that the single particle precision bound has never been experimentally surpassed with entanglement.
\end{abstract}

\maketitle

\pagestyle{plain}

By defining the standard quantum limit (SQL) as a precision of $1/\sqrt{N}$, Colombo et. al. set-up a meaningless comparison from the outset. This is not an isolated example and Colombo et. al. can be excused for following a precedent that is widely accepted in the field \cite{Meyer2001, Leibfried2004, Appel2009, Gross2010, Nagata2007, Napolitano2011, Sewell2012, Muessel2014, Kruse2016, Hosten2016a, Cox2016, Hosten2016, Braverman2019, Bao2020}. But just because everyone else is doing it, does not make it correct. Why is it incorrect? Well there are no units, no measurement time, and no mention on the number of measurements that are allowed. Furthermore the SQL is an absolute precision limit, it is neither arbitrary nor a `proportional to' or `scaling' -- bounding achievable uncertainties in such terms is meaningless.

Another definition that Colombo et. al. use pertains to measuring the spin distribution of an $N$ spin ensemble along the $S_y$ axis, where the normalised variance is $\sigma^2_y \equiv 4 (\Delta S_y)^2/N$. The SQL for a coherent spin state (CSS) with collective atomic spin $\hat{\bm{S}} = \sum \hat{\bm{s}}_i$ pointing along the $x$ axis is defined as $\sigma^2_y = 1$, i.e. $\Delta S_y = \sqrt{N}/2$. We shall call this definition the variance SQL. Why is this definition meaningless? Well, it is just a noise term which we can see by noting that $\Delta S_y$ increases with $N$. Minimising noise is not the same as improving precision. Defined in this manner the variance SQL can be manipulated to incentivise and reward poor measurements. If one simply throws aways all of the measurement data, the resulting noise with $N$ spins will be zero and the variance SQL surpassed. Thus it is not even clear what surpassing the SQL means in this definition. Colombo et. al. actually do this to some extent since they experimentally achieve $\sigma^2_{\mathrm{y}} = 0.15$ due to imperfect readout. There is no mention that this readout limitation alone prevents the SQL from being surpassed in any form. Secondly, there is nothing in the SQL that says we must use a specific atomic state that performs poorly for a given task. There are many other independent $N$ particle states that result in much lower variance along the $S_y$ axis. A CSS pointing along the $y$ axis for example can achieve $\sigma^2_y$ arbitrarily close to zero depending on experimental imperfections. In short the variance SQL does not define a measurement uncertainty since it considers only measurement noise and not the signal, and while Colombo et. al. do not explicitly claim this, many of the cited papers use such a partial and incomplete analysis to claim surpassing the SQL.

Next Colombo et. al. prepare a strongly entangled state, rotate this state by a small angle $\varphi$ around the $y$ axis, and subsequently apply a unitary which increases the angle to $m \varphi$. This is compared to a CSS with no entanglement where the state is only rotated by $\varphi$, which the authors define as the SQL. We shall call this the rotation SQL. Why is this definition meaningless? Well, one can easily surpass the rotation SQL by rotating a CSS by an angle $m \varphi$, or even rotating just a single spin by $m \varphi$. In general this would require applying the rotation unitary for a duration $m$ times longer, but as the rotation SQL places no restriction on time or allowable unitaries, any comparison is meaningless. Indeed, meaningful comparison requires defining a signal Hamiltonian which contains a parameter to be estimated and considering the measurement time required to achieve a given precision. Colombo et. al. do not demonstrate a precision better than the SQL because none of this analysis is performed, nor is it possible due to the arbitrary way the rotation SQL is defined. Again, none of the references Colombo et. al. cite demonstrate a precision for estimating a rotation angle with entangled particles that surpasses the precision bound for a single particle. Instead, the same type of arbitrary comparisons are made.

Finally Colombo et. al. report a sensitivity to magnetic fields using $N = 340$ atoms that exceeds the SQL by nearly a factor of 4. The measurement data and achieved magnetic sensitivity which would allow one to check that claim are not provided, nor is the SQL given for this task. Irrespective of meaning, the previous definitions of the SQL contain no time and have incorrect units so they can not apply. We provide it here:
\begin{equation}
\Delta B > \frac{1}{\gamma T \sqrt{N}}.
\end{equation}
Note the contrasting meaning of this definition -- with $N$ independent spins of gyromagnetic moment $\gamma$, the uncertainty $\Delta B$, in estimating the magnetic field using total measurement time $T$, can never go below this precise value -- compared to the arbitrary definitions employed previously. Based on the analysis that Colombo et. al. provide and stressing the fact that again no experiment using entanglement has ever reached a sensitivity beyond the single particle limit $\Delta B > \frac{1}{\gamma T}$, I do not expect that either of these limits have been reached here. I am offering a prize of US\$10,000 for the first demonstration of an uncertainty beyond the SQL, and an additional prize of US\$10,000 for the first demonstration of an uncertainty beyond the single particle limit when using $N$ partite entanglement. To claim the second prize, the authors need only provide evidence of an uncertainty that is within a factor of $4 \times \sqrt{340} \approx 74$ of their claimed sensitivity.

What is at stake here? 

The type of analysis that Colombo et. al. and references therein employ has led to the mistaken belief that measurements employing entanglement have achieved uncertainties that are otherwise impossible to attain. This belief has spurred the field of entanglement enhanced metrology with downstream impacts ranging from the content taught at universities to investment in efforts that leverage entanglement to build more precise devices. However, as we have shown, the field is built upon shaky foundations. Indeed my prize is motivated by work proposing that the precision obtainable with $N$ entangled particles cannot exceed the single particle limit \cite{McGuinness2021, McGuinness2021a}. As the benefits of entanglement in metrology are connected to the advantages entanglement provides in quantum information \cite{Childs2000}, such a proposition also calls into question the entire field of quantum computation.

\section{Communication with Authors and Nature Physics editor}
\noindent On July 23, 2022 a request for data was sent to the corresponding author. No data has been received to date. On August 8, 2022 this critique was sent to the authors. The corresponding author did not wish to make their response publically available.\\

\noindent On August 24, 2022 this critique, supplemented by correspondence with the authors, was submitted to \emph{Nature Physics} to be published as a Matters Arising. On Sep 6, 2022 the manuscript was rejected.

On Sep 7, 2022 I submitted the following appeal to the Nature Physics Editor:\\

Dear Leonardo,

I am surprised by your decision not to publish this Matters Arising. Not only that, I am mystified by the reasons you provided. My criticism of Vuletic and colleagues is this – they do not do what is claimed in the paper. Maybe there is some misunderstanding. Let me clarify. Vuletic and colleagues claim to use entanglement to measure some parameters with a precision that cannot be reached with the same number of independent atoms, the SQL. This is their whole paper. It is the entire abstract, the main text, all their figures and data. Without this claim, there is nothing to the paper.

So, do Vuletic and colleagues actually do this?

No.

I don’t know how to be clearer on this point. Vuletic and colleagues do not measure anything with a precision beyond the SQL. Actually, it is even worse than that, Vuletic and colleagues don’t reach a precision better than a single atom. This means that the precision actually achieved is orders of magnitude worse than claimed. This point was made clearly in the abstract of the Matters Arising - “These claims are incorrect. Colombo et. al. do not surpass the precision bound for 300 independent particles, nor do they even surpass the precision bound for a single particle.” However, I would be happy to emphasise the point further to avoid any misunderstanding.

If my criticism is valid, I can’t imagine a case that better satisfies consideration for publication as Matters Arising. If the central claims of the paper are incorrect, Vuletic’s paper should not be published anywhere, let alone Nature Physics. The only justification for not publishing this Matters Arising is that my analysis is not correct. So, is it? I direct you Vuletic’s response (submitted to you) in which he admits that they did not surpass the single atom limit:\\
\color{purple}Redacted as I have not received permission to publish this response.

\color{black} Of course, if you can’t beat the single particle limit, then you can’t beat the SQL, which Vuletic also admits:\\
\color{purple}Redacted as I have not received permission to publish this response.

\color{black}
How does this not present an interesting and timely scientific criticism and clarification? It is indeed a specific criticism pertaining to this individual paper. Furthermore, how can one square these admissions from Vuletic with the claims presented in the paper?
\begin{quotation}
``we achieve the highest metrological gain over the SQL, G = 11.8 ± 0.5 dB, that has been achieved in any (full Ramsey) interferometer to date."

``We observe a metrological gain of G = 11.8 ± 0.5 dB with $N$ = 340 ± 20 atoms in the interferometer."

``The gain G achieved with the SATIN Ramsey sequence represents a reduction by a factor of 15 in the averaging time for a desired precision."
\end{quotation}
Vuletic et. al. define the SQL in the first sentence of the abstract:
\begin{quotation}
``…with independent particles… the standard quantum limit, … limits the precision achievable in estimating unknown phase parameters."
\end{quotation}
And in the second sentence of the main text:
\begin{quotation}
``The SQL … sets the limit of precision that can be achieved with a system of $N$ independent particles."
\end{quotation}
Do Vuletic et. al. achieve the metrological gain that they claim? No. Does this gain represent a reduction by a factor of 15 in averaging time? No.

There is no grey area here, and no room for ambiguous interpretation. Vuletic and colleagues simply do not do this and they admit as such.

Regarding my criticism of other papers in the Matters Arising. If that indeed prevents publication as a Matters Arising, I can remove this commentary. However, it achieves several important goals. By not singling out Vuletic et. al. exclusively, it actually makes my criticism of their work fairer and more balanced. It also increases the interest and impact of my criticism by putting it into perspective with work of the entire community. And this clarification reduces confusion since Vuletic et. al. consistently compare their results to other published work which surpass the SQL (see e.g. Fig. 4 of the paper). Mentioning this context when critiquing Vuletic’s paper, is both beneficial and justified.

I would also like to mention that one of the claims in the paper has no supporting data, so it is impossible for anyone to check. First of all, it seems to violate the publishing standards of Nature Physics. Secondly, I have written to the authors asking to receive the data and have not received it. But most importantly, I am sure that when the authors present the data, it will confirm a sensitivity more than a factor of 100 worse than claimed.

I hope this email clarifies the matter for you, and in light of these clarifications, I strongly urge you to reconsider publishing this commentary. There does not appear to be any reason why this Matters Arising should not be published by Nature Physics.

In case you decide not to publish this Matters Arising, I would like to inform you that I will continue my efforts to ensure that the wider scientific community becomes aware of these issues. To that end, I would like your permission to include some text from your communication along with the Matters Arising manuscript posted to the Arxiv. The text that I would publish is included at the end of this email. In considering this, may I respectfully remind you of your editorial values statement: “we are committed to supporting the research enterprise by curating, enhancing and disseminating research that is rigorous, reproducible and impactful. We work to promote openness and transparency as well as the highest standards in research culture.”

I truly hope that is the case.\\

\noindent On Sep 26, 2022 the appeal was rejected.
\newpage

\end{document}